\documentstyle[aps,prd,epsf]{revtex}

\def\be{\begin{eqnarray}}
\def\ee{\end{eqnarray}}
\def\om{\omega_n}

\def\del{\partial}
\def\delmu{\partial_\mu}
\newcommand{\f}[2]{\frac{#1}{#2}}
\newcommand{\partl}[2]{\frac{\partial #1}{\partial #2}}

\begin{document}
\draft

\title{On the Convergence of the Expansion of Renormalization Group Flow
Equation}
 
\author{%
G. Papp$^{1,2,}$\thanks{on leave from HAS Research Group for Theoretical
Physics, E\"otv\"os University,  P\'azm\'any P. s. 1/A, Budapest, H-1117},
B.-J. Schaefer$^{3}$,
H.-J. Pirner$^{1,4}$ 
and 
J. Wambach$^{3}$\\[2mm]
{\small $^1$ Institut f\"ur Theoretische Physik der
Universit\"at Heidelberg,}
{\small D-69120 Heidelberg, Germany} \\
{\small $^2$ CNR,
Department of Physics, Kent State University, 44242 OH USA} \\
{\small $^3$ Institut f\"{u}r Kernphysik, TU Darmstadt, 
        D-64289 Darmstadt, Germany} \\
{\small $^4$ Max-Planck-Institut f\"ur Kernphysik, D-69029 Heidelberg, 
        Germany}
}
\date{\today}

\maketitle
\begin{abstract}
We compare and discuss the dependence of a polynomial truncation of the
effective potential used to solve exact renormalization group flow
equation for a model with fermionic interaction (linear sigma
model) with a
grid solution. The sensitivity of the results on the underlying cutoff
function is discussed. We explore the validity of the expansion method
for second and first-order phase transitions.
\end{abstract}

\section{Introduction}

Renormalization Group (RG) flow equations pioneered by Wilson and 
Ka\-danoff allow for a prediction on the behavior of a interacting
theory in different momentum regimes~\cite{wils}. 
When probing physics in the infrared (IR)
it is often useful to derive a low-energy effective theory 
by integrating over the irrelevant short-distance (high momenta) modes.  
Thus one is left with a so-called Wilsonian or low-energy effective action
distinguishing it from the one-particle-irreducible (1PI) effective action.  

In order to separate the fast-fluctuating short-distance modes from
the slowly-varying ones one has to introduce a momentum scale $k$, which 
appears naturally in the momentum cutoff regularization. The momentum 
cutoff regularization can be formulated by means of a 
blocking transformation. This procedure is achieved by introducing 
a so-called smearing or IR cutoff function $f_k$, which suppresses the 
fast-fluctuating modes of the original unblocked fields~\cite{polo}. 
In that way the fields are naturally separated into  
fast and slow mode components.  
The integration over the fast modes corresponds to the blocking
transformation on the lattice. This results in an effective action 
parameterized by the averaged blocked field at the scale $k$. The full 
1PI Feynman graphs or vertex functions, 
providing the natural tool for the study of broken symmetries in
interacting field theories, are generated by the 
renormalized effective action in the limit $k \to 0$. Thus the $k$-dependent
effective action provides a smooth interpolation between the bare action 
defined at the UV scale $\Lambda$, where no quantum fluctuations are considered
and the renormalized effective action in the IR. The RG flow pattern of the
theory is obtained by studying an infinitesimal change of the IR scale $k$,
in the effective $k$-dependent action. A nice introduction
to this topic can be found in the recent lecture~\cite{berg}.

Unfortunately, this integration step cannot be computed
in an exact way and one has to perform some approximation like the
frequently used loop expansion. Based on an one-loop expression with a
heat-kernel (proper-time) regularization we derive renormalization group 
improved flow equation and study different expansion pattern.

It is well known that a sharp momentum cutoff regularization is in
conflict with the gauge invariance~\cite{olesz,liao}.
The crucial task is to implement both the UV and the IR cutoff scales
without destroying the important symmetries. For such reasons we use the
operator cutoff regularization in Schwinger's proper-time
representation~\cite{schw} which neither depends on the dimension
nor violates any symmetry of the theory.

The paper is organized as follows: in Section~\ref{introRG} we 
define the model Lagrangian and give a general
introduction to the renormalization flow equation using the heat-kernel
regularization and different choices of the cutoff function. 
In Section~\ref{expansion} we approximate the potential by a
polynomial expansion and derive the zero temperature coupled flow
equations. Here we also present the diagrammatic interpretation of the 
equations. Temperature is introduced in Section~\ref{finiteT} and the
flow equations associated with the expansion of the potential are shown
for different cutoff functions.
In Section~\ref{result0} we present our results at zero temperature, 
comparing the results of the expansion technique at
different orders to exact grid solutions of the flow equations. 
In Section~\ref{critical} the
finite temperature calculations are presented and we discuss the
sensitivity of the critical temperature on the expansion parameter,
while in Section~\ref{exponent} the behavior of the critical
exponent, $\beta$, is studied. Finally, in Section~\ref{conclusion} we conclude
by critically assessing the feasibility of finite-density calculations using the
expansion technique. 

\section{Renormalization Group Flow Equations}
\label{introRG}
Since the intention is to study the dynamical breaking of chiral symmetry 
by matching the short-distance regime of QCD with the large-distance behavior
in the framework of evolution equations we consider an effective theory which 
incorporates simultaneously quark fields and composite mesonic fields as the 
active degrees of freedom. A suitable choice is the linear sigma model 
with $\sigma$-- and $\pi$--mesons coupled  to $N_f = 2$ constituent quarks. 
The partition function in the chiral limit  
and without external sources is given by the Euclidean-space path 
integral
\be\label{zz1}
Z & = & \int {\cal D}q {\cal D}\bar{q} 
{\cal D}\sigma  {\cal D}\vec{\pi}
 \exp\{- \int\!d^4x\, \left( {\cal L}_F + {\cal L}_B \right) \} \,,
\ee
with a fermionic part including a Yukawa coupling constant $g$, 
\be\label{fermil}
{\cal L}_F & = & \bar{q} \left( \gamma \partial +
g\left( \sigma + i\vec{\tau}\vec{\pi}\gamma_5 \right)\right) q
  \,,
\ee
and a bosonic Lagrangian part
\be\label{bosel}
{\cal L}_B & = & \frac{1}{2} \left(
(\partial_\mu \sigma)^2 + (\partial_\mu \vec{\pi})^2 \right) +
V(\sigma^2+\vec{\pi}^2) \,,
\ee
respectively. The mesonic self-interaction is parameterized by the potential
term $V$. 
At finite temperature the integration in the action along the 
temporal direction is understood on the stripe $[0,\beta]$ where $\beta$
is the inverse temperature.  
The parameters of the linear $\sigma$-model are fixed at zero
temperature as in ref.~\cite{scha}.
Because the theory is strongly coupled we cannot calculate the low-energy
theory directly. 
We assume that at an ultraviolet scale $\Lambda \approx 1$ GeV, 
the full $QCD$ dynamics reduces
to a hybrid description in terms of (massless) current quarks
and chiral bound states. The gluons
are frozen in the residual mesonic degrees of freedom, their
couplings and the {\em effective} potential $V_\Lambda$ which,
at this scale $\Lambda$, takes the following form:
\be
V_\Lambda( \vec{\phi}^2) = \frac{m^2}{2}\vec{\phi}^2
        +\frac{\lambda}{4}\vec{\phi}^4\,, \qquad 
\label{eq:pot}
\mbox{where} \quad\vec{\phi}  =  (\sigma,\vec \pi) 
\ee
is an $O(4)$-symmetric vector. Both $m^2$
and the four-boson coupling $\lambda$, are positive. 
During the course of iteratively including quantum fluctuation from a 
restricted scale range the
potential and accordingly all couplings become $k$-dependent. 
In this work we neglect the
evolution of the Yukawa coupling $g$, and set it equal a constant value 
of $3.2$ in order to reproduce the constituent quark mass $M_q \approx
300$ MeV, with the pion decay constant $f_\pi=91$ MeV, in the infrared. 


The one-loop contribution to the effective action yields in general a
non-local logarithm, which is ultraviolet divergent and needs to be
regularized.  This regularization is based on the general proper-time
representation introduced by Schwinger~\cite{schw}.  After introducing a
complete set of plane waves we are lead to the following one-loop
correction to the effective action for the bosons
\be
\label{gammab}
\Gamma^B = -\f{1}{2}\int\!d^4 x \int\limits_{0}^{\infty}\!\f{d\tau}{\tau}
\int\!\f{d^4 q}{(2\pi)^4}\, \mbox{Tr}\, 
        e^{-\tau \big(q^2 + 2i q\del + \del^2 + \left.\f{\delta^2 V}
        {\delta\phi_i\delta\phi_j}\right|_{\phi_0}\!\!\big)}
\ee
and similarly for the fermions
\be
\label{gammaf}
\Gamma^F = \f{1}{2}\int\!d^4 x \int\limits_0^\infty\!\f{d\tau}{\tau}
\int\!\f{d^4 q}{(2\pi)^4}\, \mbox{Tr}\, 
        e^{-\tau \big(q^2 + 2i q\del + \del^2 + g^2 {\phi_0}^2\big)}
\ee
where we have split the linked integration over fermions within the
bosonic integration into two parts, evaluated at the minimum $\phi_0$,
of the effective potential.  Thus the total effective action becomes a
sum of both terms $\Gamma = \Gamma^F + \Gamma^B$ each depending on
$\phi_0$. For details we refer the reader to ref.~\cite{scha}.

A prime on the potential denotes 
differentiation with respect to the fields $\phi^2$ and we elucidate our 
notation by the equation
\be
{\mbox{Tr}}\ e^{-\tau \f{\delta^2 V(\phi)}{\delta\phi_i\delta\phi_j}} = 
3 e^{-2\tau V'} + e^{-\tau\left(2 V' + 4 \phi^2 V''\right)}
 \,.
\label{trace}
\ee
The short-distance or ultraviolet divergences appear as divergences
at $\tau =0$ and the large-distance or infrared divergences as
divergences at $\tau \to \infty$.

In spite of a continuous blocking transformation procedure we modify the
above expressions by introducing a regulating function $f_k (\tau )$
into the proper-time integrand which separates the ``fast'' and ``slow''
fluctuating modes as well as respects the important symmetries of the
considered theory~\cite{Polnew}. Of course, the form of the regularized
flow equations does depend on the choice of the a priori
unknown blocking function $f_k$. In order to find an explicit form for
$f_k$ one is lead by the following conditions:
 
{\bf (1)} the modified $k$-dependent 
effective action $\Gamma_k$ should tend to the full
generating functional $\Gamma$ in the IR limit $k \to 0$, i.~e.~when the 
infrared cutoff $f_k$, is removed. Thus we require that 
$f_{k \to 0} (\tau \to \infty ) \to 1$. As a consequence in the limit 
$k \to 0$ all quantum fluctuations are taken into account.
 
{\bf (2)} in addition to the previous condition which regularizes the IR 
we set $f_{k} (\tau = 0) = 1$ for arbitrary $k$. This means that the UV regime
in the proper-time $(\tau = 0)$ is not regularized 
by this requirement.  Since we start the evolution 
at a finite (large) UV scale $\Lambda$, an UV regularization through the 
cutoff function $f_k$ is not necessary.  

{\bf (3)} a differential equation for $f_k$ can be deduced by comparing
the heat-kernel regularization to a momentum cutoff regulator in the
one-loop contribution where the one-loop trace is restricted over a
momentum slice $k \le p \le \Lambda$. It reads\footnote{A prime on $f_k$
means differentiation with respect to $k$.}:
\be
f_k' =  - (\tau k^2)^2 h(\tau k^2)
\label{eq:Diffk}
\ee
with $h(x)$ being any regular function around the origin
(cf.~\cite{scha,flor}).  Consider for example the RG improved
leading-order contribution to the one-loop potential of a scalar
one-component theory in a sharp cutoff regularization, within the
momentum shell $k \le p \le \Lambda$, (for simplicity we show it
for the second term of Eq.~(\ref{trace}), the first term can be treated
similarly.)
\be
V_k = \f{1}{2} \int\limits_{k}^\Lambda \f{d^4 p}{(2\pi)^4} \ln 
\left(p^2 + 2 V' + 4 \phi^2 V''\right)\ .
\ee
Differentiating this equation with respect to $k$ results in the flow equation
\be
k \f{\del V_k}{\del k} &=& - \f{k^4}{(4\pi)^2} \ln (k^2 + 
2 V' + 4 \phi^2 V'')\nonumber\\
& \to & \f{k^4}{(4\pi)^2} \int\limits_{0}^{\infty} \f{d \tau}{\tau }
e^{-\tau (k^2 + 2 V' + 4 \phi^2 V'')}\ ,
\ee 
where in the last line we have inserted the proper-time representation of the
logarithm.\footnote{Proper regularization is assumed but not indicated in
this formula, see e.g.~\cite{liao,norm}.}

This expression can be compared to the modified heat-kernel
regularization and yields the required differential equation of the type
(\ref{eq:Diffk})
\be\label{fk}
k f_k' = -2 (\tau k^2 )^2 e^{-\tau k^2}\ .
\ee
It is also possible to generalize this result to arbitrary $d$
dimensions with the result
\be
k f_k' = -2 (\tau k^2 )^{d/2} \f 1 {\Gamma ( d/2 )} e^{-\tau k^2}\ .
\ee
One suitable solution of Eq.~(\ref{fk}) which fulfills the proper
boundary conditions is the function 
\be
\label{eq:cutoff}
f^{(I)}_k (\tau k^2)= e^{-\tau k^2 }(1\!+\!\tau k^2)\ .
\ee
We note, that despite the evolution equation derived from a sharp momentum
cutoff and the one from the heat-kernel regularization with the smooth cutoff
function~(\ref{eq:cutoff}), have the same form, this neither means that the two
regularization methods are equivalent nor that the
choice~(\ref{eq:cutoff}) is unique. E.g. this choice for $f_k$ is
different from that in 
ref.~\cite{scha}.  In fact, it provides a slower IR convergence
as $k \to 0$. Including higher monomials of the form $\tau
k^2$ in the cutoff function $f_k$ will accelerate the IR
convergence~\cite{liao}. The choice of the cutoff function
affects the explicit form of the flow equation itself. To show this we
present our result with both the cutoff~(\ref{eq:cutoff}) (I), and the
one with higher orders,
\be
\label{eq:cutoffhi}
f^{(n)}_k (\tau k^2)= e^{-\tau k^2 }\sum_{i=0}^n \f1{i!} (\tau k^2)^i \,.
\ee
where $n=2$ corresponds to the original cutoff of ref.~\cite{scha}
$f^{(II)}_k$. We shall also examine the cutoff $f^{(III)}_k$
corresponding to $n=3$. The introduction of these higher-order
cutoff functions turns out to be even necessary in the case of 
higher-order gradient expansion in order to attain equality between the
momentum cutoff and the proper-time regularization on the level of
the wave function renormalization~\cite{olesz,liao}.

Differentiating the effective action $\Gamma_k$ or potential $V_k$ which become
$k$-dependent through the introduction of the blocking function $f_k^{(n)}$ with respect 
to the arbitrary infrared scale $k$ and after calculating the trace over inner spaces
in Eqs.~(\ref{gammab},\ref{gammaf}) yields flow equations which
incorporate only fluctuations from one-loop order and will break down
in the high temperature regime~\cite{liaop}.

This so-called ``independent-mode approximation'' can be further improved by taking
higher graphs like daisy and superdaisy diagrams~\cite{daisy} into account.
This can be accomplished by considering the interactions between the fast and
slow modes and consists of replacing the bare potential $V$ on the $rhs$
of the flow equation with the $k$-dependent potential $V_k$ 
(cf.~e.~g.~Eq.~(\ref{gammab})).

Taking this substitution of the full potential $V_k$ into account we find 
the following flow equations for the potential in $4$-dimension 
(cf.~with ref.~\cite{Janos1}) 
\be
\label{gflow0}
k\partl{V_k^{(I)}}{k}  &=& 
  \displaystyle - \f{k^4}{(4\pi)^2} \left\{
        3 \ln (1\!+\!2V_k^\prime/{k^2} ) +
          \ln (1\!+\!2V_k^\prime/{k^2}+\!4 \phi^2 V_k^{\prime\prime}/{k^2})
        \right.\nonumber \\
        &&\qquad \qquad\left.  
-\! 4N_fN_c \ln (1\!+\!g^2\phi^2/k^2) \right\} \,,
\ee
for cutoff function (I) and 
\be
\label{gflow0hi}
k\partl{V_k^{(n)}}{k} &=& 
  \displaystyle \f{k^4}{n(n\!-\!1)16\pi^2} \left\{
        \frac3{(1\!+\!2V_k^\prime/{k^2})^{n-1}} +
          \frac1{(1\!+\!2V_k^\prime/{k^2}+\!4 \phi^2
        V_k^{\prime\prime}/{k^2})^{n-1}}
  \right.\nonumber \\
  &&\qquad \qquad\left.  
-\frac{4N_fN_c}{(1 +\!g^2\phi^2/k^2)^{n-1}} \right\} \,.
\ee
for the other cutoff functions 
($n=2$ corresponds to case (II) and $n=3$ to case (III)), respectively.
Using the smooth cutoff function for case (I) $f_k^{(I)}$
yields a Renormalization Group equation with a characteristic
logarithmic structure very similar to the Wegner-Houghton equation
with a sharp-cutoff~\cite{wegn}.  

The first term on the $rhs$ of Eqs.~(\ref{gflow0}--\ref{gflow0hi}) 
represents the flow equation contribution from the 
three massless pions\footnote{$V_k^\prime$ vanishes at the minimum}, 
the middle term the sigma-meson contribution 
with mass squared term 
$2V_k^{\prime}\!+\!4 \phi^2V_k^{\prime\prime}$ and the last term the
$N_f N_c$ quark contribution.\footnote{particle -- antiparticle and spin $1/2$ 
yields the factor 4.}

As a result of the choice of the cutoff function $f_k^{(n)}$ for $n>1$
(cf.~ref.~\cite{scha}) the flow equation contributions have the typical form
$1/(k^2 + \mbox{mass}^2)$ and are sometimes called ``threshold
functions''~\cite{wett}. This structure controls the decoupling of the
massive modes in the infrared.
In the case of cutoff function (I) however, the ``threshold functions'' are
represented by logarithms.  Nevertheless, the appearance of the mass
terms serve here the same purpose, namely to regularize the evolution
equations. 

At first glance the structure of the nonperturbative flow
Eqs.~(\ref{gflow0}--\ref{gflow0hi}) looks different. Eq.~(\ref{gflow0})
is the one-loop resummation of the perturbative expansion and can be
rewritten by expanding the logarithm in the non-Gaussian
pieces~\cite{Alex} of the potential. Comparing the resulting series with
the expansion of the other flow Eqs.~(\ref{gflow0hi}) one finds similar
terms relating Eqs.~(\ref{gflow0}--\ref{gflow0hi}) to each other.

On the other hand one can expand the effective action in powers of momenta. In
coordinate space this expansion takes the form~\cite{Janos2,Bonanno} 
(gradient expansion) 
\be
\Gamma [\phi_0 ] = \int d^4 x \left\{ - V [\phi_0] + 
        \f12 Z[\phi_0] (\delmu \phi_0)^2 + 
        Y[\phi_0] (\delmu \phi_0)^4 + \cdots \right\}\ .
\ee
The lowest-order approximation of this expansion is the local potential
approximation (LPA) which consists of considering a constant background
field, setting $Z =1$ and neglecting all higher coefficients
functions $Y, \cdots$. Now the truncation becomes apparent and
one sees the restriction to the
effective potential neglecting the influence of the anomalous dimension
$\eta$ defined by $-k\f {\partial \ln Z}{\partial k}$. 
A more general investigation of an $O(N)$-symmetric potential including
anomalous dimensions within this approach will be published elsewhere~\cite{bohr}.  


\section{Expansion of the Potential}
\label{expansion}

In the following we introduce a short-hand notation $\rho \equiv
\phi^2$, and expand the potential $V_k$ at the scale $k$ around its
minimum to study the stability of such an expansion as the function of
order $M$. A similar analysis involving only mesonic fields was done
recently in~\cite{Aoki}.

Explicitly, for the symmetric phase ($\rho_0 = 0$) up to order
$M$,
\be
  V_k^{(s)} = \sum_{n=0}^M \f{1}{(2n)!} a_{2n}(k) \rho^n
\ee 
and similarly for the broken phase around the new $k$-dependent
nontrivial vacuum ($\rho_0 \neq 0$),
\be
  V_k^{(b)} = b_0+\sum_{n=2}^M \f{1}{(2n)!} b_{2n}(k)
\left(\rho\!-\!\rho_0(k)\right)^n
        \,.
\ee
The second coefficient $b_2$ of this expansion defines the minimum
$\rho_0$ of the potential and therefore vanishes.
In the local potential approximation the expansion coefficients of
the potential correspond to the $2n$-point proper vertices evaluated
at zero momenta which we have denoted by $a_{2n}$ for the symmetric phase. 
Now substituting the upper
expansions on both sides of the flow equation (\ref{gflow0}) we can deduce
a coupled system of flow equations for the proper vertices.    
For the symmetric phase (evaluated at $\rho_0 = 0$)  
we obtain the following set of coupled equations for the first four
couplings $a_0$, $a_2$, $a_4$ and $a_6$ with the cutoff function (I),
\be
\label{syma0}
k \partl{a_0}{k} &=& \displaystyle \f{k^4}{4\pi^2} \ln{
        \f{k^2}{k^2\!+\!a_2}} \\
\label{syma2}
k \partl{a_2}{k}  &=& \displaystyle - \f{k^4}{8\pi^2} \left\{
\f {a_4}{k^2\!+\!a_2} - 4 N_c N_f \f{g^2}{k^2} \right\}\\
\label{syma4}
k \partl{a_4}{k}  &=& \displaystyle - \f{k^4}{4\pi^2} \left\{
\f25 \f{a_6}{k^2\!+\!a_2} -\f {a_4^2}{(k^2\!+\!a_2)^2}
+12 N_c N_f \f {g^4}{k^4} \right\}\\
\label{syma6}
k \partl{a_6}{k}  &=& \displaystyle - \f{k^4}{(4\pi)^2} \left\{
\f{10}7 \f{a_8}{k^2\!+\!a_2} - 18\f {a_4 a_6}{(k^2\!+\!a_2)^2}
+ \f{100}3 \f {a_4^3}{(k^2\!+\!a_2)^3}  
-960 N_c N_f \f {g^6}{k^6} \right\}\,. 
\ee
Note, that 
the flow equation of the first coupling $a_0$, has a contribution from the
quarks in the symmetric phase through $a_2$.
Similar equations may be obtained using the cutoff functions (II) and (III).

The integration of the high-momenta degrees of freedom results in new
one-loop flow contributions to the couplings of the effective Wilsonian
Lagrangian. During the evolution towards the IR all possible
higher-dimension composite interactions are generated which are even
partially nonrenormalizable. However, the contributions of these
nonrenormalizable vertices are under control, since the scales involved
during the evolution are much smaller than the corresponding cutoff
values. Hence, these operators have a negligible effect
on the physics at infrared scales.

Equations~(\ref{syma2}--\ref{syma6}) are easily identified 
as the differential form of the truncated Dyson-Schwinger equations at
vanishing external momenta.
The denominator $k^2 + a_2$ plays the role of the bosonic propagator
with the boson self-energy squared $a_2(k)$ and 
Eq.~(\ref{syma2}) corresponds to the bosonic self-energy equation at
vanishing external momenta,
\be
\partl{\Sigma^2_k (0)}{k} \ \propto \  \quad
\raisebox{-6mm}{\epsfbox{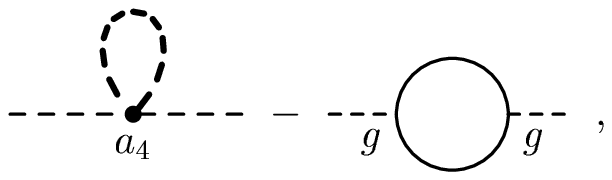}}
\label{dys-schwing}
\ee
while Eq.~(\ref{syma4}) is the Bethe-Salpeter equation for the vertex
function at vanishing external momenta,
\be
\partl{\Gamma^{(4)}_k(0)}{k} \ \propto \  \quad
\raisebox{-10mm}{\epsfbox{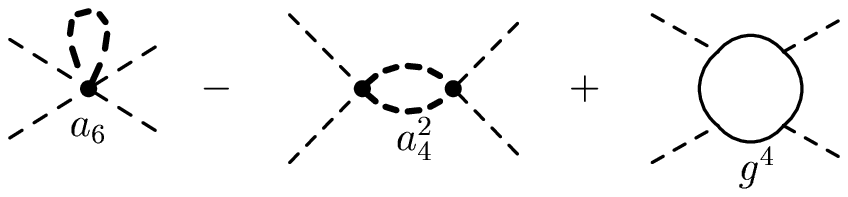}}
\label{beth-salp}
\ee
without the integration in the loop-momentum variable $k$.
In the figure the single external lines represents the low-momenta field
(which do not appear in the equations), while the inner lines are dressed
by the high-momenta fluctuations in one-loop corrections down to scale $k$.
Each of Eqs.~(\ref{syma2}--\ref{syma6}) symbolizes the sum of all one-loop
diagrams with a fixed number $2n$ of external legs~\cite{Bonanno}.  The
analogous fermionic inverse propagator is associated with a
massless quark in the symmetric phase, and is proportional to $k$.

For the broken phase we obtain the following flow equation structure
expanding around the nontrivial minimum $\rho_0$,
\be
\label{brokb0}
k \partl{b_0}{k} &=& - \f{k^4}{(4\pi)^2} \left\{
        \log(1+b_4\rho_0/3k^2)-4 N_c N_f\log(1+g^2\rho_0/k^2) \right\}\\
\label{brokrho0}
k \partl{\rho_0}{k} &=& \f{k^4}{(4\pi)^2 b_4} \left\{
        \f{6b_4}{k^2}+\f25 \f{15 b_4+\rho_0 b_6}{k^2+b_4\rho_0/3}
                -\f{48 g^2 N_c N_f}{k^2+g^2\rho_0} \right\} \\
\label{brokb4}
k \partl{b_4}{k}  &=& \displaystyle - \f{k^4}{(4\pi)^2} \left\{
\f 3 5 \f{b_6}{k^2} - \f{b_4^2}{k^4} -\f 1 {75}
\f{(15 b_4 + \rho_0 b_6)^2}{(k^2 + \rho_0 b_4 / 3)^2}
+ \f 1 {35} \f{35 b_6 + \rho_0 b_8}{(k^2 + \rho_0 b_4 / 3)}
\right.\nonumber\\
&\vdots&\left. \qquad\qquad +\f{48 g^4 N_c N_f}{(k^2 + g^2 \rho_0)^2} \right\}
+ \f{b_6}{10} k \partl{\rho_0}k \,.
\ee  
We omit here the explicit form of the higher flow equations because the
result is lengthy but straightforward. If one defines the mass of the
$\sigma$-particle as $m^2_{\sigma} = \rho_0 b_4/3$ then one can easily 
identify the massive $\sigma$-propagators which must be distinguished
from the massless pion propagators in the chiral limit. On the other
hand the quarks are now massive $m_q^2 = g^2 \rho_0$ in the broken
phase. Furthermore one sees nicely the matching of both equations
(\ref{syma4}) and (\ref{brokb4}) during the course of evolution with
respect to the scale $k$ because, at the chiral symmetry breaking scale,
$a_2$ and $\rho_0$ tend both to zero.  This feature also applies to
the flow equations for higher couplings $b_6$ etc..

Once again the diagrammatic interpretation of Eq.~(\ref{brokb4}) is
straightforward,
\be
\partl{\Gamma^{(4)}_{k}(0)}{k} &\propto& \quad
\raisebox{-9mm}{\epsfbox{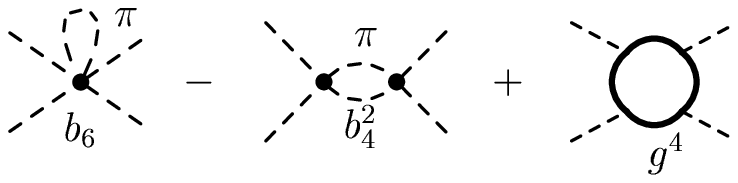}}
\nonumber\\[0mm]
&-& \quad
\raisebox{-7mm}{\epsfbox{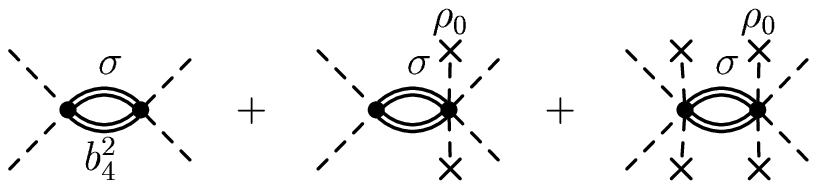}}
\label{beth-salpb}
\\[2mm]
&+& \quad
\raisebox{-7mm}{\epsfbox{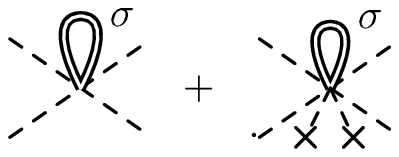}}
\nonumber
\ee
The crosses on the graphs symbolize the vacuum expectation value $\rho_0$. 
The massive $\sigma$-propagator is represented by a double line, the 
massive $\pi$-propagator as a dashed line and the massive
quark-propagator as a thick solid line. Below 
several graphs the order of the corresponding couplings is indicated.

We note, that the form of the higher-order cutoff functions
$f^n_k$, in the heat-kernel regularization corresponds to adding $n-1$ ghost
fields~\cite{liao} to the loop integrals i.e. additional graphs appear in 
Eqs.~(\ref{dys-schwing},\ref{beth-salp},\ref{beth-salpb}).


\section{Finite-Temperature Flow Equation}
\label{finiteT}
It is straightforward to generalize the above approach to finite temperature.
In order to match the four--dimensional theory at zero temperature with
the effectively purely three--dimensional behavior at the
critical temperature it is more convenient to use the imaginary time
approach (Matsubara formalism). 
This means that we replace the four--dimensional momentum integration in 
Eqs.~(\ref{gammab}--\ref{gammaf}) by a three--dimensional integration and
introduce a discrete Matsubara summation with the corresponding frequencies
for bosons $\om = 2n\pi T$, and fermions $\nu_n = (2n\!+\!1)\pi T$, 
in the zero-momentum component.

Furthermore, an additional length scale $\beta$, the inverse
temperature, is introduced. This incorporates additional thermal
fluctuations beyond the already existing quantum ones. As shown in
the appendix of ref.~\cite{scha} it is possible to split the thermal
fluctuations into a zero temperature contribution plus a finite
temperature contribution using the so-called generalized
$\Theta$-function transformation. However, in the present work we
decided to solve the Matsubara sums numerically and do not apply the
$\Theta$-function transformation.

Thus at finite temperature the scale $k$  serves as a
generalized IR cutoff for a combination of the three-dimensional momenta
and Matsubara frequencies. A careful investigation of how the two types
of modes are controlled by different choices of the blocking function
can be found in ref.~\cite{liao98}.  
In the present work we use a cutoff function family with a
four-dimensional momentum variable, related to the smeared version of type
(3) in ref.~\cite{liao98}.

Using the cutoff function (I) -- Eq.~(\ref{eq:cutoff}) -- we obtain the
following finite-temperature (dimensional) flow equation for the full
potential
\be
k \partl{V_k^{(I)}}{k} = &
\displaystyle \f{k^3}{8\pi} 
        T \sum\limits_{n= -\infty}^\infty \left\{ \f3{[
                1\!+\!(\om^2\!+\!2V^\prime_k)/k^2 ]^{1/2}}\!+\!
        \f1{[ 
                1\!+\!(\om^2\!+\!2V^\prime_k\!+\!4\phi^2 
                V^{\prime\prime}_k)/k^2 ]^{1/2}}
        \right. \nonumber \\
        & \left. \displaystyle
        -\ \f{4N_cN_f}{[
                1\!+\!(\nu_n^2\!+\!g^2\phi^2)/k^2 ]^{1/2}} \right\} 
\label{gflowI}
\ee
and
\be
k \partl{V_k^{(n)}}{k} =  \displaystyle \f{k^3}{8\pi} T\ 
        \f{(2n\!-\!3)!!}{n!\  2^{n\!-\!1}}
         \sum\limits_{n= -\infty}^\infty \left\{ \f3{[
                1\!+\!(\om^2\!+\!2V^\prime_k)/k^2 ]^{n\!-\!1/2}}\ +
        \qquad\qquad\right. \nonumber \\
        \qquad\qquad \left. \displaystyle
        \f1{[ 
                1\!+\!(\om^2\!+\!2V^\prime_k\!+\!4\phi^2 
                V^{\prime\prime}_k)/k^2 ]^{n\!-\!1/2}}
        -\ \f{4N_cN_f}{[
                1\!+\!(\nu_n^2\!+\!g^2\phi^2)/k^2 ]^{n\!-\!1/2}} \right\}
\label{gflowhi}
\ee
with $(2n\!-\!3)!! = 1 \cdot 3 \ldots \cdot (2n\!-\!3)$
for cutoff II ($n=2$) and III ($n=3$). 
Due to the three-dimensional momentum integration 
fractional powers arise in the threshold functions. 
In appendix A of ref.~\cite{scha} a relation for the low-temperature 
limit of the Matsubara sums 
to the four-dimensional integrals can be found. This guarantees the correct 
matching of the finite-temperature equation to the zero temperature one 
since in the limit $T \to 0$ the two become identical.


\section{Results at $T=0$.}
\label{result0}
At zero temperature the general flow
equations~(\ref{gflowI}--\ref{gflowhi}) reduce to
the flow equations~(\ref{gflow0}--\ref{gflow0hi}) which can be proved
analytically using the low-temperature limit of the Matsubara sums
as described in the appendix of ref.~\cite{scha}.

We have solved numerically the flow equations~(\ref{gflowI}--\ref{gflowhi})
with the cutoff functions (I)-(III) both on a grid and using the set of
coupled equations on the expansion coefficients~(\ref{syma0}--\ref{syma6}) 
and ~(\ref{brokb0}--\ref{brokb4}) and compared the results. In the grid
calculation the potential was discretized by 80 points in the range 
$0<\phi^2<0.05$ GeV$^2$ and the set of 80 coupled differential equations
were solved with a Runge Kutta method.

We start the evolution in the symmetric phase deep in the ultraviolet
region of QCD at $\Lambda = 1.2$ GeV and fix our parameters similarly to
ref.~\cite{scha} for cutoff function (II), to get at the end of the
evolution the physical value of the pion decay constant, and the chiral
symmetry breaking scale in the range of $0.8-1$ GeV.

In table~\ref{tab1para} we list the used parameterization for the cutoff
functions (I), (II) and (III).  In the course of the evolution
towards zero with respect to $k$ we encounter the transition to chiral
symmetry breaking at a finite scale $k_{\chi SB}$, where we switch to
the equations for the broken phase. The chiral symmetry breaking scale
depends on the underlying cutoff function and is also listed in
table~\ref{tab1para}.  The required initial values are chosen in such a way
as to fix the pion decay constant $f_\pi$, at the end of the evolution in
the infrared.

\begin{figure}[thbp]
\centerline{\epsfxsize=0.85\textwidth \epsfbox{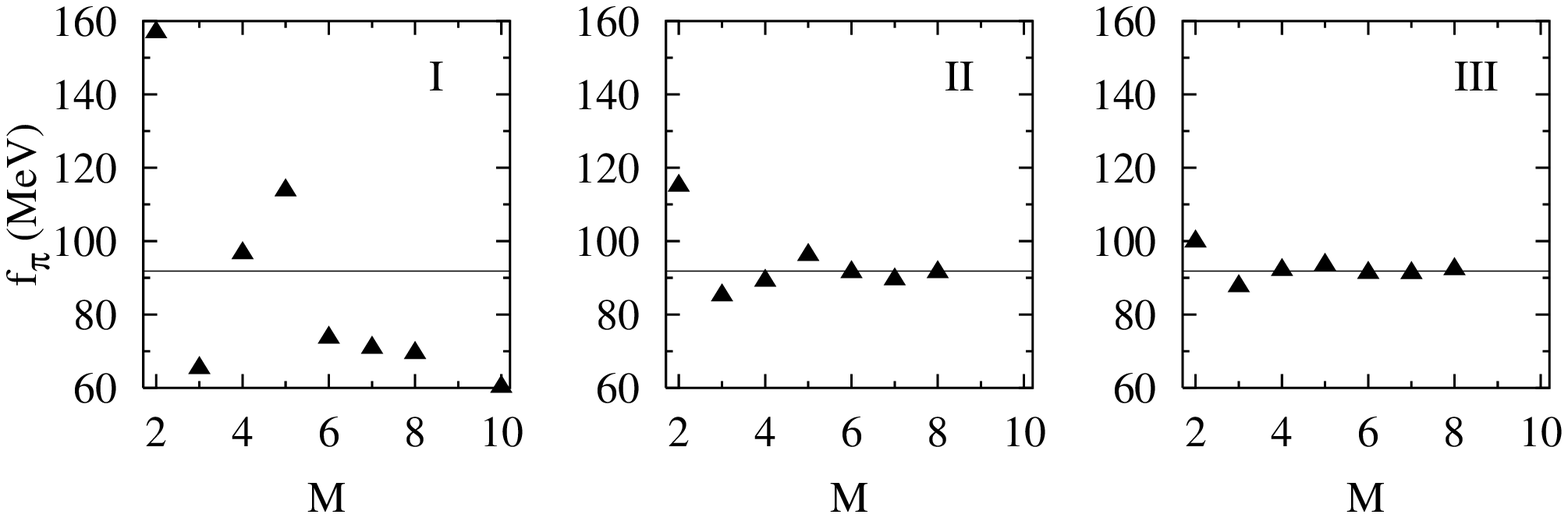}
}
\caption{Dependence of the pion decay constant $f_\pi$, on the
order of expansion for different cutoff functions. 
The solid lines represent the grid values.}
\label{figpiI}
\end{figure}

\begin{figure}[thbp]
\centerline{\epsfxsize=0.85\textwidth \epsfbox{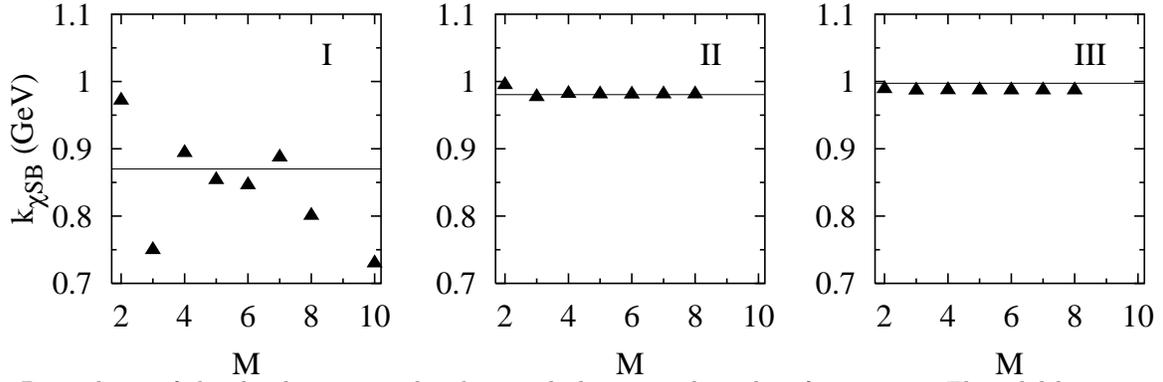}
}
\caption{Dependence of the chiral symmetry breaking scale $k_{\chi SB}$,
on the order of expansion. The solid lines represent the grid values.}
\label{figkthI}
\end{figure}

In Fig.~\ref{figpiI} we compare the convergence of the expanded
equations as a function of the order of the expansion. The solid lines
represent the grid value of $f_\pi$ and the triangles are the ones
related to the solution of the expanded equations for the cutoff I
(left), II (middle) and III (right). A similar plot for the chiral symmetry
breaking scale $k_{\chi SB}$ is presented in Fig.~\ref{figkthI}.

The lesson we learn from the figure is that for the cutoff (I) the convergence
is very poor, actually the higher-order calculations are getting even
worse, while for the higher-order cutoffs the convergence is improving
rapidly with the order of the cutoff $n$. The lowest-order $M=2$,
truncation overestimates $f_\pi$ by $70\%$ for cutoff (I), by
$25\%$ for cutoff (II) and only by $8\%$ for cutoff (III). Cutoff (I) shows
unstable behavior when increasing the order of expansion, while cutoff (II) 
gives already a good approximation to the full (not expanded)
calculation at $M=6$ and cutoff (III) at $M=4$. However, for all cutoffs,
taking higher-order polynomials, the global shape 
of the potential becomes worse going away from the minimum, that is the
convergence radius of the approximation becomes smaller and smaller around the
minimum. This is demonstrated by calculating the infrared limit of the
expansion coefficients.

As mentioned above, we fix the parameters of the quartic
potential~(\ref{eq:pot}) at the momentum scale $\Lambda=1.2$ GeV,
setting the initial conditions for the evolution equations as $a_0=0$,
$a_2=m^2$, $a_4=6 \lambda$ and $a_{2i}=0$ for $i\ge3$.
At the chiral symmetry breaking scale $k_{\chi SB}$, the mass parameter
$a_2$, becomes negative and the minimum of the potential shifts to some
finite $\rho_0$ value. Below this momentum scale
we use Eqs.~(\ref{brokb0}--\ref{brokb4}) for the broken phase.
Studying the structure of these equations and higher orders in the $k\to0$
limit, one finds
\be
\label{explim4}
\partl{b_4}{k} = \f{36}{\pi^2} \f {b_4^2}{k} \qquad
\mbox{yielding} \quad b_4^{(III)} = -\f{\pi^2}{36 \ln{k}}
\ee
and
\be
\label{explim6}
\partl{b_6}{k} = -\f{768}{5\pi^2} \f {b_4^3}{k^3} \qquad
\mbox{yielding} \quad b_6^{(III)} = -\f{2\pi^4}{1215\, k^2 \ln^3{k}} \,.
\ee
Generally, 
\be
\label{explim8}
b_{2i}^{(III)} \propto \f1{k^{2i\!-\!4}\ln^i{k}} \,.
\ee
This result is also supported by the numerics. Since the
asymptotic structure is similar for all cutoff functions, we get similar 
results,
\be
b_4^{(I)} = -\f{16\pi^2}{\ln{k}} \qquad \mbox{and} \quad
b_4^{(II)} = -\f{16\pi^2}{\ln{k}}
\ee
for the cutoff I and II.

The expansion coefficients for $2i\ge6$ diverge in the infrared limit,
hence making the expansion of the potential meaningless outside the
minimum. 


\section{Results at $T\neq 0$.}
\label{critical}

\begin{figure}[thbp]
\centerline{\epsfxsize=0.85\textwidth \epsfbox{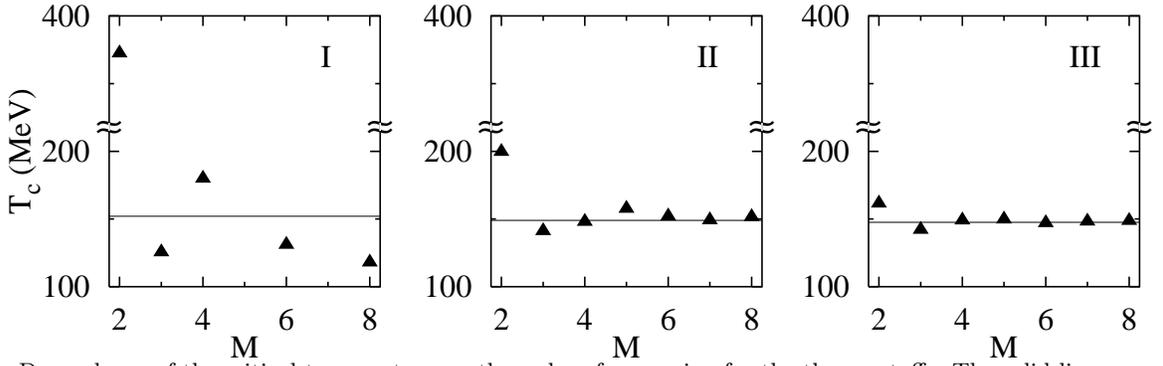}}
\caption{Dependence of the critical temperature on the order of
expansion for the three cutoffs. The solid lines 
represent the grid values. Note the cut on the y-axis. See text.}
\label{fig2}
\end{figure}

In Fig.~\ref{fig2} we present the critical temperature
as the function of the order of polynomial expansion for the same
initial parameter set as for zero temperature. Once again there is no
convergence obtained for cutoff (I), however, cutoffs (II) and (III) behave
properly. The grid value of the critical temperature is $T_c\approx
150$ MeV  for all three cutoffs. For cutoff (II), $M=2$ overshoots the
asymptotic value by $25\%$ and stabilizes at $M\ge6$. Cutoff (III) is much
more stable, deviating only by $9\%$ from the grid value at $M=2$ and
stabilizing at $M=4$.

\begin{figure}[thbp]
\centerline{\epsfxsize=0.85\textwidth \epsfbox{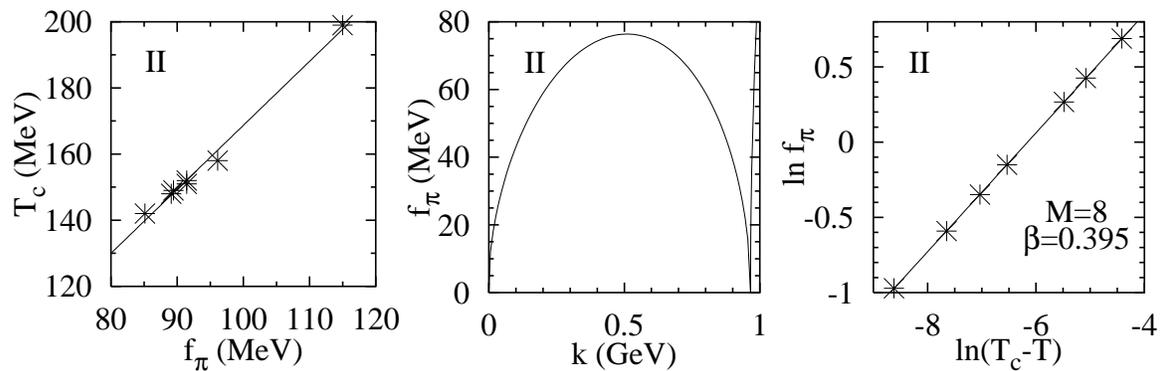}}
\caption{Correlation of the pion decay constant at zero temperature and
the critical temperature (left) for cutoff (II). Evolution of the pion
decay constant with the momentum scale at the critical temperature
(middle) and calculation of the critical exponent for cutoff (II),
$M=8$. See text.} 
\label{fig3}
\end{figure}

The evolution of the minimum of the potential at the critical
temperature shows the same interesting and counterintuitive pattern
already discussed in~\cite{scha}: decreasing the momentum scale 
starting from the symmetric phase in the deep ultraviolet one enters the
broken phase due to the presence of the quarks
and the minimum of the potential $\phi_k = f_\pi(k)$, first increases up
to $k\approx 0.5$ GeV, then decreases back to zero (see left side of
Fig.~\ref{fig3}). Such a behavior is confirmed by the calculations made on
the grid, the shape of $f_\pi$ as a function of $k$ can well be fitted by
an ellipsis. Above the critical temperature lowering the momentum scale
hence there are two transitions, one from the initial symmetric phase to
a broken one at $k_{\chi SB}$, and at small $k$ another transition is
driving the system 
back to the symmetric phase. This indicates that the infrared modes now
restore the symmetry. Such a behavior was also observed in other works
with a different cutoff family~\cite{wett,wettp}.

There is an interesting relation that can be found between the value of the
pion decay constant at zero temperature and the critical temperature for
the same initial parameter set but different expansion orders. The
obtained values follow nicely a linear fit (see left part of
Fig.~\ref{fig3}), in accordance with a slope parameter of 2 found
in~\cite{bilic} in the case of the NJL-model.

The asymptotic behavior of the expansion coefficients $b_{2i}$, may be
studied analytically in the $k\to 0$ limit of the broken phase. Since in
this limit the relevant quantity $T/k$ is approaching infinity, the
Matsubara sum may be replaced by its lowest-order contribution. Keeping
the leading-order terms in the momentum scale $k$, we arrive at
\be
  b_{2i} \propto \f{k^2}{(Tk)^{i\!-\!1}}
\ee
for $i\ge1$ and up to $T\le T_c$. Once again the coefficients are
divergent for $i>3$ shrinking the convergence radius of the expansion to
one point.


\section{Critical exponents}
\label{exponent}

\begin{figure}[thbp]
\centerline{\epsfxsize=0.85\textwidth \epsfbox{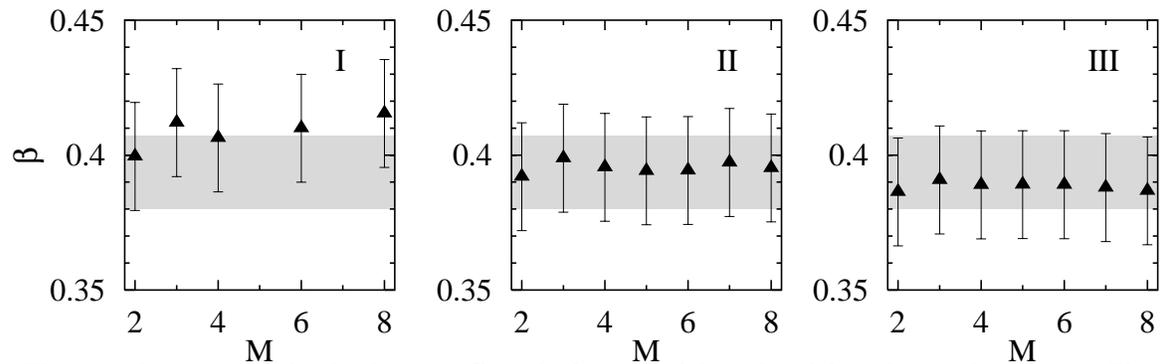}}
\caption{The critical exponent $\beta$, for the three cutoffs as the
function of the order of the polynomial expansion. The shadowed region
represents the $O(4)$ value from different calculations. See text.}
\label{critI}
\end{figure}

Our numerical results show a second-order phase transition in the order
parameter $f_\pi$, that is the order parameter vanishes continuously
with a critical exponent $\beta$, $f_\pi \sim (T_c-T)^{\beta}$. A typical
plot, for $M=8$, is presented in Fig.~\ref{fig3} (right), in the scaling
window of the model. The
critical exponent $\beta$, is only slighlty sensitive to the order of the
polynomial expansion and cutoff used and agrees with the
three-dimensional scalar $O(4)$-Heisenberg model value,
$\beta\approx0.4$. For comparison lattice Monte Carlo simulations yield
$\beta = 0.3836(46)$ \cite{kana}, $4-\epsilon$ expansion
$\beta = 0.38(1)$ \cite{epsilon}
and the average action approach 
$\beta = 0.407$ \cite{wett}. 
In fig.~\ref{critI} we demonstrate the order dependence $M$ of the
critical exponent for the different cutoff functions (I)-(III) and
compared our values with the one in the literature: the shadowed region
represents the value region obtained in different calculations. Despite
the fact that
the convergence of the pion decay constant and critical temperature was
poor (even absent) in the case of the cutoff function (I), the critical
exponent $\beta$, is still quite well in agreement with the $O(4)$-model
value. For the higher cutoff functions (II) and (III)
the convergence becomes more and more stable, however, a systematic
decrease with the order $n$, of the cutoff functions is observed in the
direction of the value obtained by the $4-\epsilon$ expansion and the MC
calculations. This decrease persists further at $M=4$, $n=4$ approaching
closer to the $4-\epsilon$ expansion value.

It is also interesting to compare the size of the scaling window for
different cutoff functions. At $M=4$ we found that the window starts at
the values of $(T_c-T)/T_c=$ 17\%, 7\%, 2.7\% and 0.8\% for cutoffs
(I) through (IV), respectively. This indicates a widening of the scaling
regime for lower monomial cutoff functions.


\section{Conclusions}
\label{conclusion}
We have studied a model Lagrangian of quarks and mesons with renormalization
group equations based on an one-loop expansion within the local
potential approximation, for different cutoff functions.
Expanding the meson potential in polynomials $\phi^{(2M)}$, of degree
$M$ around its minimum, we found that the lowest order
cutoff function (I), related to the Wegner-Houghton equation with a
sharp cutoff has serious
convergence problems. Increasing the order of the cutoff function
applied, the convergence in the order of the potential expansion $M$,
increases rapidly, already with the next cutoff
function (II), the expansion in the potential is within 25\% 
for $M=2$ ($\phi^4$ theory). In order to get results closer to the full
potential (grid) calculation however, one has to take higher corrections
into account. 
The analysis of the critical behavior in ref.~\cite{wett} has shown that
the ``magnetization'' $\phi_0$, is related to the external source $J$,
(the current quark mass in the present model) as $\phi_0 \sim
J^{1/\delta}$ with the critical index $\delta\approx4.8$. Hence the
effective potential has the dependence $V_{eff} \sim \phi^{1+\delta}$
around the minimum. This explains the necessity to have at least terms
with $M=3$ corresponding to $\phi^6$ in the effective potential near
$T_c$.
Our study shows that for cutoff function (II) 
with $M=6$ expansion
yields 
stable results and close to the full grid simulation
for several global quantities (such as the pion decay constant, the chiral
symmetry breaking scale $k_{\chi SB}$,
and the critical temperature $T_c$). However, for local
quantities, such as the potential itself, increasing order gives less
and less reliable shape (the convergence radius of the expansion defines
a smaller and smaller interval around the minimum), deviating from the
full grid calculation. This means, that studies involving first-order
phase transition (as encountered at finite baryon density~\cite{wettmu,meye}) 
cannot be
treated in such a model up to the infrared scale, since the two minima
are separated further than the convergence radius and cannot be
calculated
reliably. The problem may be addressed using full grid calculations
or other expansion basis may be considered, however, for finite
density such a basis should include non-analytical functions.

\acknowledgments
One of the authors (HJP) would like to thank J\"urgen Berges to bring
the subject into his attention. G.P. thanks Michael Strickland for early
discussions. 
This work was supported in part by DOE grant DE-FG02-86ER40251,
NSF grant NSF-Phy98-00978, Hungarian grant OTKA F026622 and German grant
IKDA 15/99.


\vfill
\eject


\begin{table}[htb!]
\caption{The initial parameters $m$ and $\lambda$ at the scale $k=\Lambda$ 
for
different cutoff functions (I)--(III) and the resulting pion decay
constant $f_\pi$, at the scale $k\to0$. The chiral symmetry breaking scale
$k_{\chi SB}$, is also given.}
\label{tab1para}
\begin{tabular}[htb!]{c|d|d|d|d}
cutoff function & $m$ {[GeV]} & $\lambda$ & $f_\pi$ [MeV] & $k_{\chi
SB}$ [GeV] \\ \hline
(I)             & 0.4   & 51.5      &  91.85        & 0.87 \\ \hline
(II)            & 0.4         & 30.0        &  91.86        & 0.98 \\ \hline
(III)           & 0.4   & 16.0      &  91.86        & 1.00  
\end{tabular} 
\end{table}


\begin{thebibliography}{99}
\bibitem{wils}
L.~P.~Kadanoff,
        Physica {\bf 2}, 263 (1966);
K.~G.~Wilson,
        Phys. Rev. {\bf B4}, 3174 and 3184 (1971); 
K.~G.~Wilson and J.~G.~Kogut,
        Phys. Rev. {\bf 12}, 75 (1974).

\bibitem{polo}
S.-B.~Liao and J.~Polonyi,
        Annals of Physics {\bf 222}, 122 (1993).

\bibitem{berg}
J.~Berges,
        {\tt hep-ph/9902419}.

\bibitem{olesz}
M.~Oleszczuk,
	Z. Phys. {\bf C64}, 533 (1994).

\bibitem{liao}
S.-B. Liao,
	Phys. Rev. {\bf D53}, 2020 (1996).

\bibitem{schw}
J.~Schwinger,
        Phys.~Rev.~{\bf D13}, 3224 (1976).

\bibitem{scha} 
B.-J. Schaefer and H.-J. Pirner,
        Nucl.Phys. {\bf A627}, 481 (1997);
B.-J. Schaefer and H.-J. Pirner,
        {\tt nucl-th/9903003}.

\bibitem{Polnew}
S.-B. Liao, J. Polonyi and M. Strickland,
        {\tt hep-th/9905206}.

\bibitem{flor}
R.~Floreanini and R.~Percacci,
        Phys. Lett. {\bf B356}, 205 (1995).

\bibitem{norm}
S.~W.~Hawking, 
	Commun. Math. Phys. {\bf 55}, 133 (1977).

\bibitem{liaop} 
S.-B. Liao and J. Polonyi,
        Phys. Rev. {\bf D51}, 4474 (1995).

\bibitem{daisy} 
L.~Dolan and R.~Jackiw,
        Phys. Rev. {\bf D9}, 3357 (1974).

\bibitem{Janos1}
J. Alexandre, J. Polonyi,
        {\tt hep-th/9902144}.

\bibitem{wegn}
F.~J.~Wegner and A.~Houghton,
        Phys. Rev. {\bf A8}, 401 (1973);
F.~J.~Wegner, 
in {\em Phase Transitions and Critical Phenomena}, vol.~6 eds. C. Domb
and M. Greene (Academic Press, 1976).

\bibitem{wett}
C. Wetterich,
        Phys. Lett. {\bf B301}, 90 (1993);
C. Wetterich and N. Tetradis,
        Int. J. Mod. Phys. {\bf A9}, 4029 (1994);
N. Tetradis and C. Wetterich,
	Nucl. Phys. {\bf B422}, 541 (1994);
D.-U. Jungnickel and C. Wetterich,
        Phys. Rev. {\bf D53}, 5142 (1996);
J. Berges, D.-U. Jungnickel and C. Wetterich,
        Phys. Rev. {\bf D59}, 034010 (1999).

\bibitem{epsilon}
G. A. Baker, B. G. Nickel and D. I. Meiron,
        Phys. Rev. {\bf B17}, 1365 (1978);
J. Zinn-Justin,
        {\em Quantum Field Theory and Critical Phenomena} (Oxford
        University Press, 1990) Chapter 25 and references therein. 

\bibitem{kana}
K. Kanaya and S. Kaya, 
        Phys. Rev. {\bf D51}, 2404 (1995).

\bibitem{Alex}
J. Alexandre, V. Branchina and J. Polonyi,
        Phys. Rev. {\bf D58}, 016002 (1998).

\bibitem{Janos2}
T. R. Morris,
        Phys. Lett. {\bf B334}, 355 (1994);
T. R. Morris, M. D. Turner,
        Nucl. Phys. {\bf B509}, 637 (1998);

\bibitem{Bonanno}
A. Bonanno, V. Branchina, H. Mohrbach and D. Zappal\`a,
        {\tt hep-th/9903173}.

\bibitem{bohr}
O.~Bohr, B.-J. Schaefer and J.~Wambach;
        in preparation.

\bibitem{Aoki}
K.I. Aoki, K. Morikawa, W. Souma, J.I Sumi and H. Terao,
        Prog. Theor. Phys. {\bf 99}, 451 (1998);
J.O Andersen, M. Strickland,
        {\tt cond-mat/9811096}, and references therein.

\bibitem{liao98} 
S.-B. Liao and M. Strickland,
        Nucl. Phys. {\bf B532}, 753 (1998).

\bibitem{wettp}
D.-U. Jungnickel, 
	private communications

\bibitem{bilic}
N. Bili\'c, J. Cleymans and M.D. Scadron,
        Int. J. Mod. Phys. {\bf A10}, 1169 (1995).

\bibitem{wettmu}
J. Berges, D.-U. Jungnickel and C. Wetterich,
        {\tt hep-ph/9811347} and {\tt hep-ph/9811387}.

\bibitem{meye}
J. Meyer, G. Papp, H.-J. Pirner and T. Kunihiro,
        {\tt nucl-th/9908019}.

\end{thebibliography}
\end{document}